\documentclass[aps,prd,showkeys,showpacs,preprintnumbers,amsmath,amssymb,superscriptaddress,floatfix,nofootinbib]{revtex4}

\usepackage{epsfig}
\usepackage{graphics}
\usepackage{latexsym}
\usepackage{amsmath}
\usepackage{amssymb}
\usepackage{rotating}
\usepackage{subfigure}
\usepackage{bm}
\usepackage{multirow}
\usepackage{color}
\usepackage{amsfonts}
\usepackage{amsmath}
\usepackage{mathtools}
\usepackage{float}
\usepackage{epstopdf}
\usepackage[misc]{ifsym}
\usepackage{enumitem}

\begin{document}

\title{The opportunity to find $\bar{d}^\ast(2380)$ in the $\Upsilon(nS)$ decay}

\author{Chao-Yi L\"{u}}\email{lvcy@ihep.ac.cn}
\affiliation{Institute of High Energy Physics, CAS, Beijing 100049, China}
\affiliation{Department of Physics, Zhengzhou University, Zhengzhou, Henan 450001, China}
\author{Ping Wang}\email{pwang4@ihep.ac.cn}
\affiliation{Institute of High Energy Physics, CAS, Beijing 100049, China}
\affiliation{Theoretical Physics Center for Science Facilities, CAS, Beijing 100049,
China}
\author{Yu-Bing Dong}
\affiliation{Institute of High Energy Physics, CAS, Beijing 100049, China}
\affiliation{Theoretical Physics Center for Science Facilities, CAS, Beijing 100049,
China}
\affiliation{School of Physical Sciences, University of Chinese Academy of Sciences, Beijing, 101408 China}
\author{Peng-Nian Shen}
\affiliation{Institute of High Energy Physics, CAS, Beijing 100049, China}
\affiliation{Theoretical Physics Center for Science Facilities, CAS, Beijing 100049,
China}
\author{Zong-Ye Zhang}
\affiliation{Institute of High Energy Physics, CAS, Beijing 100049, China}
\affiliation{Theoretical Physics Center for Science Facilities, CAS, Beijing 100049,
China}
\affiliation{School of Physical Sciences, University of Chinese Academy of Sciences, Beijing, 101408 China}

\begin{abstract}
$d^\ast(2380)$ was observed by WASA-at-COSY collaborations in the nuclear reaction recently. Its particularly narrow width may indicate the new QCD-allowed hadronic structure. To further confirm the existence of this peculiar particle in a totally different kind of reaction, we study the opportunity for searching $\bar{d}^\ast$ in the $\Upsilon(nS)$ (with n=1, 2, 3) decay. As a phenomenological study, our framework is based on SU(3) chiral quark model. By virtue of the unitarity of $S$-matrix and crossing symmetry, we study the imaginary part of the forward scattering amplitudes between $\Upsilon(nS)$ and $d^\ast$. The scattering process is mainly governed by the quark-meson interaction. We examine both the pseudoscalar and vector meson contribution in the intermediate state. Hopefully, our results show that it's quite possible to find $\bar{d}^\ast$ in this decay mode in the future.
\end{abstract}

\maketitle

\section{Introduction}\label{section1}
The research about six quark system started more than 50 years ago when Dyson and Xuong proposed $\Delta\Delta$ structure from a simple group classification without considering dynamical effects \cite{Dyson:1964xwa}. At that time, compared with the blank in experiment data, the relevant theoretical work went in advance. Recently, more and more XYZ zoos and kinds of exotic resonances are discovered. Tetraquarks, pentaquarks, molecular states, hybrid states, glueball, ..., all these hypothetical structures are allowed by quantum chromodynamics (QCD), which makes the research of these peculiar particles much attractive. Along with the flourish of this field these years, many efforts were put into experiments. The data was accumulated increasingly. As a consequence, a dibaryon-like exotic state came out from the theory, present in our front.

The signature of such a resonance was reported by WASA-at-COSY collaborations when they studied the ABC effect in $pn\rightarrow d\pi^0\pi^0$, $pn\rightarrow d\pi^+\pi^-$ reactions \cite{P. Adlarson:2011,M. Bashkanov:2009}. The further confirmation was done in series of reactions, for example, $pn\rightarrow pp\pi^-\pi^0$, $pn\rightarrow pn\pi^0\pi^0$, $pd\rightarrow {}^3He\pi^-\pi^+$, $pd\rightarrow {}^3He\pi^0\pi^0$, etc \cite{S.
Keleta:2009,Adlarson:2012au,Adlarson:2013usl,Adlarson:2014tcn,Adlarson:2014xmp,Adlarson:2014pxj}. After some analysis, it is shown that the quantum number of this resonance is $I(J^P)=0(3^+)$ \cite{Clement:2016vnl}. For the description of the observables, the collaboration obtained the mass and width of the s-channel resonance being $M=$ 2380 MeV, $\Gamma=$ 70 MeV, using the Breit-Wigner ansatz for the resonance amplitude.
Facing its mass and extraordinarily narrow width, the conventional $\Delta\Delta$ picture doesn't work well. But on the other hand, the data clearly show a new dibaryon-like structure from the Dalitz plot \cite{M. Bashkanov:2009,Adlarson:2014tcn}. All these messages together imply that there may exist a new structure in this particle.

The news from experimentalists spread around quickly and kindled the enthusiasm in theorists in a moment. Many hypotheses were proposed and various methods were employed to uncover the property of this particle \cite{Gal:2013dca,Gal:2014zia,Chen:2014vha,Huang:2013nba,Lu:2018wsr,Dong:2017mio,M. Bashkanov:2013,Dong:2018emq,Dong:2017geu,Dong:2016rva,Dong:2015cxa,Lu:2017uey}. Among these theories, Gal and Garcilazo examined this structure by solving $\pi N \Delta$ Faddeev three-body equations \cite{Gal:2013dca,Gal:2014zia}.
The other major structural model which can reasonably explain the data was supported by Zhang {\it et al.}. In their theory, the particle is the $\Delta\Delta$ system couples with the hidden color channel $(CC)$. The earliest work was done in 1999 \cite{YZYS99}. After the recent experimental observations, Huang {\it et al.} followed this ideal, completed
a dynamical coupled-channels study in the framework of the resonating group method (RGM) \cite{Huang:2015nja}.

Up to now, Only WASA-at-COSY collaborations claim the existence of this particle. And $d^\ast(2380)$ has only shown up in the nuclear reaction process. Obviously, apart from the discussions on its structure, being observed in just one kind of reaction is far from enough. Further examination of its existence in totally different mechanism is extremely necessary. Relevant theoretical research on this topic is still scanty. For this reason, it motives us to explore the opportunity to search this particle in a different experiment. From the data given by Particle Data Group (PDG), we find that anti-deuteron can be produced in the $\Upsilon(nS)$ decay \cite{pdg}. Since $\bar{d}^\ast$ is a six anti-quark system similar to anti-deuteron, and the mass of $\Upsilon(nS)$ is larger than two times of that of $\bar{d}^\ast$, the phase space is large enough for $\Upsilon(nS)$ decaying into $\bar{d}^\ast$ $d^\ast$ pair or $\bar{d}^\ast$ plus other hadrons. It naturally inspires us that maybe anti-$d^\ast$ can show up in this decay mode, just like anti-deuteron. If this decay channel can be measured, it implies the existence of $\bar{d}^\ast$ and consequently $d^\ast$.

In this paper, we firstly evaluate the unknown parameters by fitting the partial widths of process $\Upsilon(nS)\rightarrow \bar{d}+X$. Then for $\bar{d}^\ast$ production case, we replace the wave function of deuteron with that of $d^\ast$ obtained by RGM and the parameters are chosen in a relatively broad range based on the deuteron case
to estimate the differential widths.

The paper is organized as follows. In Sect. \ref{section2}, we present the brief formulism including both interaction and wave functions. The numerical results are discussed in Sect. \ref{section3}. Finally, Sect. \ref{section4} is a summary.

\section{Brief Formulism}\label{section2}

\begin{figure}[!htbp]
\centering
\includegraphics [width=6cm, height=5cm]{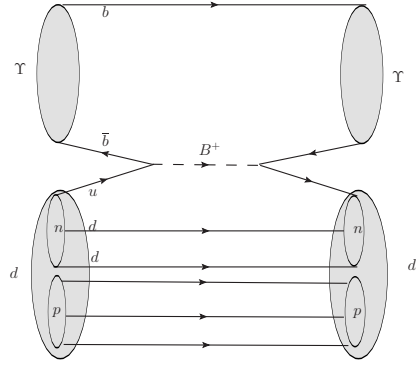}
\caption{The Feynman diagram of $\Upsilon+ d\rightarrow \Upsilon+d$ forward scattering.} \label{Fig1}
\end{figure}

The differential decay widths of $\Upsilon(nS)\rightarrow\bar{d}+X$ is given by the formula
\begin{widetext}
\begin{equation}\label{eq1}
\begin{split}
\textrm{d}\Gamma=&\frac{1}{2m_\Upsilon}(\frac{\textrm{d}^{}\vec{p}_{\bar{d}}}{(2\pi)^{3}}\frac{1}{2E_{\bar{d}}})\sum_{n-1}(\prod^{n-1}_{f}\int\frac{\textrm{d}^{}\vec{p}_{X_i}}{(2\pi)^{3}}\frac{1}{2E_{X_i}})  \left|\mathcal{M}(\Upsilon\rightarrow \bar{d}+X)\right|^2
(2 \pi)^{4}\delta^{4}(p_{\Upsilon}-p_{\bar{d}}-\sum p_{X_i}),
\end{split}
\end{equation}
\end{widetext}
where we clearly separate the part related to $\bar{d}$ from others.
The crossing symmetry tells us that
\begin{equation}\label{eq2}
\mathcal{M}(\Upsilon\rightarrow \bar{d}+X)
=\mathcal{M}(\Upsilon+d\rightarrow X),
\end{equation}
with $p_d=-p_{\bar{d}}$.
As a consequence of the unitarity of the $S$-matrix, when we insert $S=1+iT$ into $S^\dag S=1$, we have
\begin{equation}\label{eq3}
-i(T-T^\dag)=T^\dag T.
\end{equation}
Sandwich the left and right hand side of Eq.~$\left(\ref{eq3}\right)$ between the same initial and final states of $\Upsilon(nS)$ and deuteron, and insert a complete set of intermediate states between $T^\dag$ and $T$ in the right hand side to count any possible physical processes, then we can obtain the equation related to our process
\begin{widetext}
\begin{equation}\label{eq4}
2\textrm{Im}\mathcal{M}(\Upsilon+d\rightarrow\Upsilon+d)
=\sum_{n}(\prod^{n}_{i=1}\int\frac{\textrm{d}^{}\vec{p}_{X_i}}{(2\pi)^{3}}\frac{1}{2E_{X_i}})\times\left|\mathcal{M}(\Upsilon+d\rightarrow X)\right|^2\delta^{4}(p_{\Upsilon}+p_{d}-\sum_{i}p_{X_i}),
\end{equation}
\end{widetext}
which together with crossing symmetry relates the imaginary part of the forward scattering $\Upsilon(nS)+d\rightarrow\Upsilon(nS)+d$ to the differential decay width of $\Upsilon(nS)\rightarrow\bar{d}+X$.
So the original formula for decay width can be converted into
\begin{equation}\label{eq:wid}
\textrm{d}\Gamma= \frac{1}{m_\Upsilon}(\frac{\textrm{d}^{}\vec{p}_{d}}{(2\pi)^{3}}\frac{1}{2E_{d}})\times \mathrm{Im}\mathcal{M}(\Upsilon+d\rightarrow \Upsilon+d),
\end{equation}
which is a consequence of optical theorem.
For the forward scattering processes $\Upsilon(nS)+d\rightarrow \Upsilon(nS)+d$, we assume the dominant mechanism is the fusion of the $\bar{b}$-quark and the light quark, and the produced $B$ meson plays the role of s-channel intermediate state. This is because the quarks in $\Upsilon$ and deuteron are in different types, there is no meson exchange between them. Therefore, we adopt the meson fusion mechanism in constituent quark model to calculate the elementary process $\bar{b}+u(d)\rightarrow\bar{b}+u(d)$. The Feynman diagram for the $B^+$ meson case is shown in Fig.~\ref{Fig1}.

In our model, the Lagrangian for the pseudoscalar meson can be written as
\begin{equation}
{\cal L}=ig_{ps}\bar{\psi}_{\bar{b}}\gamma_5\psi_qB,
\end{equation}
where $g_{ps}$ is the coupling constant. $\bar{\psi}_{\bar{b}}$, $\psi_q$ and $B$ describe the fields of the light flavor quark, $\bar{b}$-quark and B meson, respectively.
The Lagrangian relevant to vector meson $B^\ast$ is expressed as
\begin{equation}
{\cal L}=g_{v}\bar{\psi}_{\bar{b}}\gamma_\mu\psi_qB^{\ast\mu}+\frac{f_{v}}{4M}\bar{\psi}_{\bar{b}}\sigma_{\mu\nu}\psi_q(\partial^\mu B^{\ast\nu}-\partial^\nu B^{\ast\mu}),
\end{equation}
where $g_{v}$ and $f_{v}$ denote the coupling constants for vector and tensor interaction, respectively. $M$ is a mass scale, which can be taken as the mass of $\Lambda_b$.
With the Lagrangian, the pseudoscalar meson contribution for this process can be written as
\begin{eqnarray} \label{eq:ampl}
\begin{split}
{\cal M}=\dfrac{\big|g_{ps}\big|^2}{q^2-m_B^2+im_B\Gamma}\Psi^\ast_\Upsilon\Psi^\ast_d\bar{u}(p_u)\gamma_5
v(p_{\bar{b}})\bar{v}(p_{\bar{b}})
\gamma_5u(p_u)\Psi_d\Psi_\Upsilon,
\end{split}
\end{eqnarray}
where $\Psi_\Upsilon$ and $\Psi_d$ are the wave functions of $\Upsilon(nS)$ and deuteron, respectively. $u(p_u)$, $\bar{u}(p_u)$, $\bar{v}(p_{\bar{b}})$ and $v(p_{\bar{b}})$ represent the spinors of the $u(d)$-quark and $\bar{b}$-quark in the initial and final states, respectively. $q$ is the momentum of the $B$ meson. As for the contribution of the vector meson, the amplitude is expressed as
\begin{widetext}
\begin{eqnarray} \label{eq:ampl}
\begin{split}
{\cal M}=&\big|g_{v}\big|^2\Psi^\ast_\Upsilon\Psi^\ast_d\bar{u}(p_u)\gamma_\mu
v(p_{\bar{b}})\dfrac{g^{\mu\nu}-q^\mu q^\nu/q^2}{q^2-m_{B^\ast}^2+im_{B^\ast}\Gamma}\bar{v}(p_{\bar{b}})\gamma_\nu u(p_u)\Psi_d\Psi_\Upsilon\\
&+\frac{\big|f_{v}\big|^2}{(2M)^2}\Psi^\ast_\Upsilon\Psi^\ast_d\bar{u}(p_u)\sigma_{\mu\alpha}q^\mu
v(p_{\bar{b}})\dfrac{g^{\alpha\beta}-q^\alpha q^\beta/q^2}{q^2-m_{B^\ast}^{2}+im_{B^\ast}\Gamma}\bar{v}(p_{\bar{b}})\sigma_{\nu\beta}q^\nu u(p_u)\Psi_d\Psi_\Upsilon\\
&+\frac{i g_{v}f_{v}^\dag}{2M}\Psi^\ast_\Upsilon\Psi^\ast_d\bar{u}(p_u)\sigma_{\mu\alpha}q^\mu
v(p_{\bar{b}})\dfrac{g^{\alpha\beta}-q^\alpha q^\beta/q^2}{q^2-m_{B^\ast}^{2}+im_{B^\ast}\Gamma}\bar{v}(p_{\bar{b}})\gamma_{\beta} u(p_u)\Psi_d\Psi_\Upsilon\\
&-\frac{if_{v} g_{v}^\dag}{2M}\Psi^\ast_\Upsilon\Psi^\ast_d\bar{u}(p_u)\gamma_{\alpha}
v(p_{\bar{b}})\dfrac{g^{\alpha\beta}-q^\alpha q^\beta/q^2}{q^2-m_{B^\ast}^{2}+im_{B^\ast}\Gamma}\bar{v}(p_{\bar{b}})\sigma_{\nu\beta}q^\nu u(p_u)\Psi_d\Psi_\Upsilon,
\end{split}
\end{eqnarray}
\end{widetext}
The formular for the process $\Upsilon(nS)\rightarrow\bar{d}^\ast+X$ is the same as that for deuteron case except the wave function changed from derteron to $d^\ast$.

The realistic wave function of $\Upsilon(nS)$ is obtained by solving Schr\"{o}dinger equation with the well known Cornell potential $V(r)=-\frac{4\alpha}{3r}+\lambda r+c$ \cite{Y. Ding:1999,Y. Lu:2016,Y. Lu:2017}. The wave functions of deuteron and $d^\ast$ are taken from the previous works \cite{Huang:2015nja}. In the quark degrees of freedom, the wave function of deuteron can be written as
\begin{eqnarray} \label{eq:wfd}
\begin{split}
\Psi_{d}=&{\cal A}\left[\phi_{p}\phi_{n}
\eta_{NN}^{l=0}(\vec{R})\zeta_{d}\right]_{(SI)=(10)},
\end{split}
\end{eqnarray}
where ${\cal A}$ is a total anti-symmetrization operator. $\phi_p$ and $\phi_n$ are the internal wave functions of proton and neutron, respectively. $\eta^{l=0}_{NN}$ denotes the relative wave function with zero orbital momentum, which is calculated dynamically with the chiral SU(3) constituent quark model. $\zeta_{d}$ represents the required wave function for spin-isospin. The internal wave function of nucleon has the form of
\begin{eqnarray}
\phi_{n(p)}=\dfrac{1}{\sqrt{2}}\left[\chi_\rho\psi_\rho
       +\chi_\lambda\psi_\lambda\right]\Phi(\vec{\rho},\vec{\lambda}),
\label{eq:wfn}
\end{eqnarray}
where $\chi_\rho$ ($\chi_\lambda$) and $\psi_\rho$ ($\psi_\lambda$) stands for the symmetric (anti-symmetric) wave functions in the spin and isospin spaces, $\Phi(\vec{\rho},\vec{\lambda})$ is responsible for spatial part with $\rho$ and $\lambda$ the Jacobi coordinates. Similar to deuteron, the description of $d^\ast$ is taken as
\begin{eqnarray}\label{eq:wfd*}
\begin{split}
\Psi_{d^\ast}&={\cal A}\big[\phi_\Delta(\vec{\rho}_1,\vec{\lambda}_1)
\phi_\Delta(\vec{\rho}_2,\vec{\lambda}_2)~\eta^{l=0}_{\Delta\Delta}(\vec{R})
\zeta_{\Delta\Delta}\\
&+\phi_{C_8}(\vec{\rho}_1,\vec{\lambda}_1)
\phi_{C_8}(\vec{\rho}_2,\vec{\lambda}_2)\eta^{l=0}_{C_8C_8}(\vec{R})\zeta_{C_8C_8}\big]_{(SI)=(30)},
\end{split}
\end{eqnarray}
with $\phi_\Delta$, $\phi_{C_8}$ being the inner wave functions of $\Delta$ (color-singlet) and $C_8$ (color-octet) clusters. $\eta^{l=0}_{\Delta\Delta}$ and $\eta^{l=0}_{C_8C_8}$ denote the $S$-wave relative wave functions between $\Delta\Delta$ and $C_8C_8$ clusters (the $D$-wave components are negligibly small and not considered here). $\zeta_{\Delta\Delta}$ and $\zeta_{C_8C_8}$ stand for the spin-isospin wave functions in the hadronic degrees of freedom in the corresponding channels. It should be emphasized that when we combine the spin-isospin
wave functions, the quantum numbers for the $C_8$-cluster and the $\Delta$-cluster in deuteron are different. For the color-octet, we
have $S = 3/2$; $I = 1/2$; $C = (1~1)$, while for the color-singlet, $S = 3/2$; $I = 3/2$; $C = (0~0)$, where $S$, $I$ and $C$ denote the spin, isospin and color of the six-quark system, respectively.
To reduce the tedious calculation and make the two components orthogonal to each other, the relative wave function $\eta^{l=0}_{\Delta\Delta(C_8C_8)}$ is projected to $\chi^{\text{eff},l=0}_{\Delta\Delta(C_8C_8)}$ by using the so-called channel wave function of $\Delta\Delta$ ($C_8C_8$) cluster
\begin{equation}
\chi^{\text{eff},l=0}_{\Delta\Delta(C_8C_8)}(\vec{R})=\left<\phi_{\Delta(C_8)}(\vec{\rho}_1,\vec{\lambda}_1)
\phi_{\Delta(C_8)}(\vec{\rho}_2,\vec{\lambda}_2)\Big|\Psi_{d^\ast}\right>.
\end{equation}
Similarly, the wave function of $d^\ast$ is reassembled as
\begin{eqnarray}
\begin{split}
\Psi_{d^\ast}\cong& \big[\phi_\Delta(\vec{\rho}_1,\vec{\lambda}_1)
\phi_\Delta(\vec{\rho}_2,\vec{\lambda}_2)\chi^{\text{eff},l=0}_{\Delta\Delta}(\vec{R})
\zeta_{\Delta\Delta}\\
&+\phi_{C_8}(\vec{\rho}_1,\vec{\lambda}_1)
\phi_{C_8}(\vec{\rho}_2,\vec{\lambda}_2)\chi^{\text{eff},l=0}_{C_8C_8}(\vec{R})\zeta_{C_8C_8}\big]_{(SI)=(30)}.
\end{split}
\end{eqnarray}
It should be particularly mentioned that the resultant effective wave function maintains all the effect of total anti-symmetrization of the wave function of $d^\ast$. After projecting onto the physical base, the effective wave function is further expanded by the sum of four Gaussian functions
\begin{eqnarray}
\label{eq:chwf}
\begin{split}
\chi^{\text{eff},l=0}_{\Delta\Delta(C_8C_8)}(\vec{R})=\sum_{i=1}^{4}c_i
\textrm{exp}\left(-\dfrac{\vec{R}^2}{2b_i^2}\right).
\end{split}
\end{eqnarray}
More details about the wave functions can be found in Ref.~\cite{Huang:2015nja}.

Since the width of $B$ meson is very narrow, the Breit-Wigner form of the propagator can be approximated as a $\delta$-function
\begin{equation}\label{eq:propa}
\dfrac{1}{q^2-m_B^2+im_B\Gamma}\simeq\frac{1}{q^2-m_B^2+i\epsilon}\rightarrow
-2\pi i\delta(q^2-m_B^2),
\end{equation}
when we only focus on the imaginary part of the amplitude. In the $B^\ast$ case, we also have the similar approximation.

The remained work is to simplify the spinors with the numerator in propagator to the kinematic variables. The sum over quark spins can be easily performed using the completeness relations. After contracting the indices and evaluating the trace, we arrive at the desired expressions. For the intermediate pesudoscalar meson, the spinor part is expressed as
\begin{equation}\label{eq:pse}
S_{PS}=\big|g_{ps}\big|^2\left(-4E_{\bar{b}}E_q-4m_{\bar{b}}m_q\right),
\end{equation}
where $E_{\bar{b}}$ and $E_q$ are energies of the $\bar{b}$-quark and light quark. For the vector meson, the vector coupling gives
\begin{equation}\label{eq:vec}
S_V=\big|g_v\big|^2\left(-4E_{\bar{b}}E_q-12m_{\bar{b}}m_q-\frac{8}{m_B^2}\left(m_{\bar{b}}^2 m_q^2+(m_{\bar{b}}^2+ m_q^2)E_{\bar{b}}E_q+E_{\bar{b}}^2E_q^2+(\vec{p}_{\bar{b}}\cdot \vec{p}_q)^2\right)\right),
\end{equation}
where $\vec{p}_{\bar{b}}$ and $\vec{p}_q$ are momenta of the $\bar{b}$-quark and light quark, respectively.
The expression of the spinnor part for the tensor coupling is written as
\begin{equation}\label{eq:ten}
S_T=\frac{\big|f_v\big|^2}{M^2}\left(-4\left(m_{\bar{b}}^2 m_q^2+(m_{\bar{b}}^2+ m_q^2)E_{\bar{b}}E_q+E_{\bar{b}}^2E_q^2+(\vec{p}_q\cdot \vec{p}_{\bar{b}})^2\right)+m_{\bar{b}}m_qm_B^2\right).
\end{equation}
The last term is for the interference between vector and tensor coupling, expressed as
\begin{equation}\label{eq:int}
S_{VT}=\frac{6(g_v f_v^\dag+g_v^\dag f_v)}{M}\left(m_{\bar{b}}(m_q^2+E_{\bar{b}}E_q)+m_q(m_{\bar{b}}^2+E_{\bar{b}}E_q)\right).
\end{equation}
The terms odd in $\vec{p}_{\bar{b}}$ do not contribute to the integral and they are not shown in the above expression.

The momentum of $d$ or $d^\ast$ is related to the particles generated from the $\Upsilon$ decay.
Because of the lack of the information on $X$ in final state, we simply insert a phenomenological form factor to describe the momentum distribution of deuteron and $d^\ast$. For example, for deuteron case
the form factor $F(p_d)$ is assumed to be
\begin{equation}\label{eq:dis}
F(p_d) ={\cal N} \text{exp}\left [-\frac{(p_d-p_0^d)^2}{(\Lambda^{d})^2}\right ],
\end{equation}
where ${\cal N}$ is the normalization factor. $p_0^d$ and $\Lambda^d$ will be determined by the experimental momentum distribution of anti-deuteron in $\Upsilon(nS)$ decay.

\section{Numerical results}\label{section3}
\begin{figure*}[!htbp]
\centering
\includegraphics [width=5.5cm, height=4.5cm]{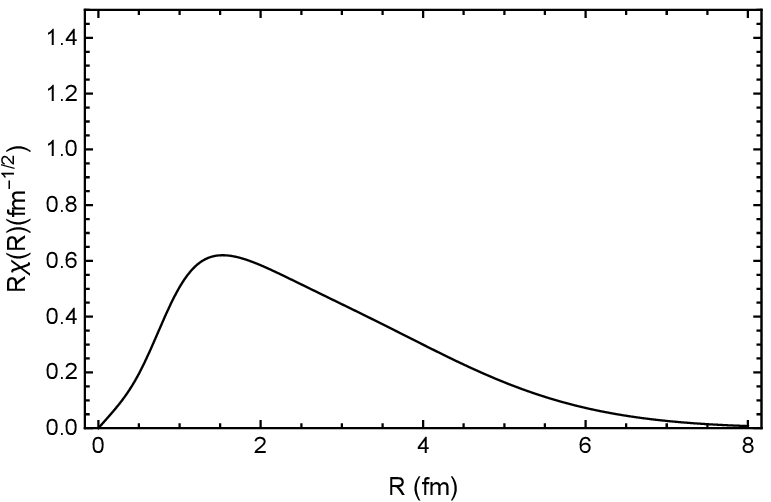}
{\hskip 0.25cm}
\includegraphics [width=5.5cm, height=4.5cm]{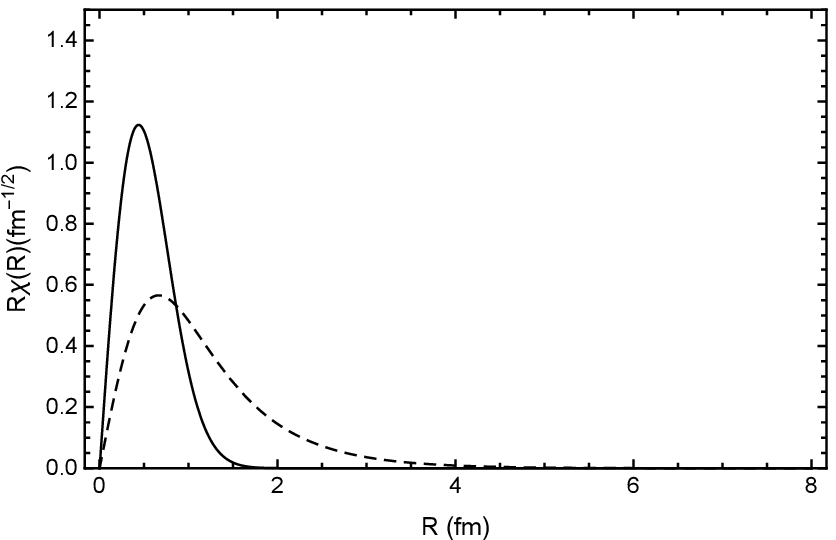}
\caption{Channel wave functions in the $S$-wave by using the
extended chiral SU(3) quark model. The solid curve in the left
diagram describes the wave function for deuteron, and the dashed and
solid curves in right depict the wave functions for the $\Delta\Delta$ and
$C_8C_8$ components of $d^\ast(2380)$, respectively.} \label{fig:wf}
\end{figure*}

In this section, we will show the numerical results. Firstly, in Fig.~\ref{fig:wf}, we present the $S$-wave channel wave function of deuteron and $d^\ast$ obtained from RGM.
As we can see from these curves, the positions of the peaks are located at 1.5 fm, 0.8 fm, 0.5 fm for deuteron, $\Delta\Delta$ and $C_8C_8$ clusters in $d^\ast$, respectively, which means $\Delta\Delta+CC$ components make $d^\ast$ more compact than deuteron, obviously.

In our numerical calculation, the masses of light quark and $\bar{b}$ quark are chosen as $m_q=0.313$ GeV and $m_{\bar{b}}=4.96$
GeV. $B$ meson mass is chosen to be 5.28 GeV. The three coupling constants $g_{ps}$, $g_v$ and $f_v$
have not been determined. Fortunately, from the observation of the momentum dependence of Eqs.~(\ref{eq:pse})$-$(\ref{eq:int}), we find that the pseudoscalar and vector meson contributions
are close to each other.
This is because with the restriction of the $\delta$-function in Eq. (\ref{eq:propa}), even the upper limits of $p_{\bar{b}}$ and $p_q$ are much smaller than $m_{\bar{b}}$ and $m_q$, respectively. As a result, the matrix elements of $\bar{b}+u(d)\rightarrow\bar{b}+u(d)$ are not sensitive to $p_{\bar{b}}$ and $p_q$. In other words, although the analytic expressions for pseudoscalar and vector meson are different, their contribution can be summed up and the total contribution is determined by one effective coupling constant $g^{\text{eff}}$.

\begin{figure*}[!htbp]
\centering
\includegraphics [width=5.5cm, height=4.5cm]{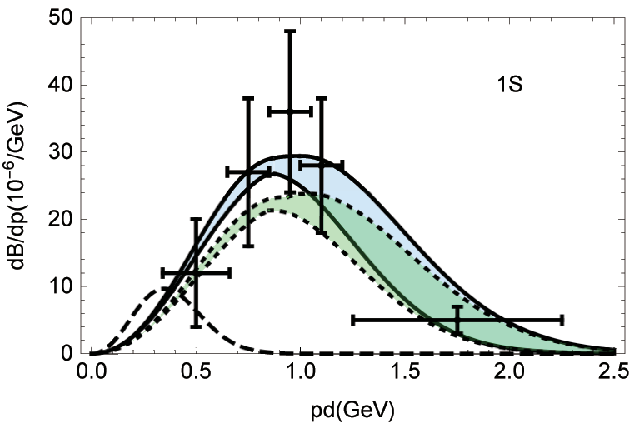}
{\hskip 0.25cm}
\includegraphics [width=5.5cm, height=4.5cm]{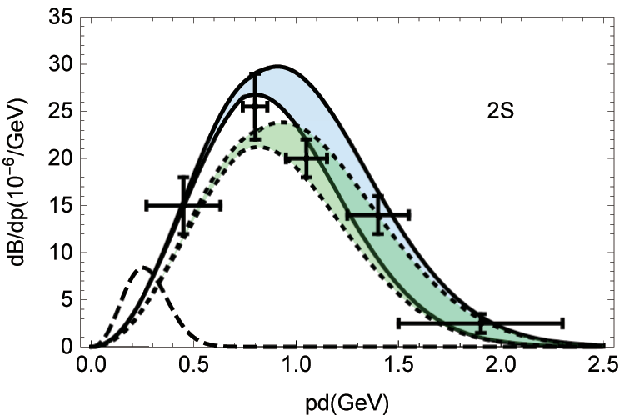}
{\hskip 0.25cm}
\includegraphics [width=5.5cm, height=4.5cm]{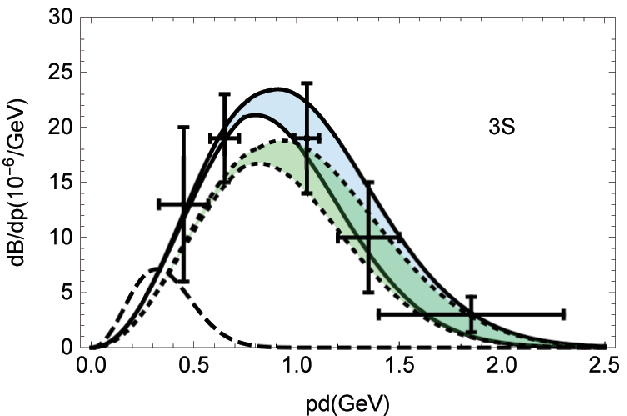}
\caption{The differential decay widths for the process
$\Upsilon(nS) \rightarrow \bar{d} (\bar{d}^\ast) + X$. The figures from left to right are for
decay of $\Upsilon(1S)$, $\Upsilon(2S)$ and $\Upsilon(3S)$, respectively. The dots with error bars are the experimental data \cite{CLEO,babar}. The bands formed by solid lines are the fits for $\bar{d}$ and the bands formed by dotted lines are predictions for $\bar{d}^\ast$. The dashed line on the left in each figure is for the prediction at lower limit.} \label{fig:width}
\end{figure*}

Therefore, with the good approximation, we only need to determine three unknown paramaters $g^{\text{eff}}$, $p_0^d$ and $\Lambda^d$. They are fixed by the experimenal momentum distribution of $\bar{d}$ in the $\Upsilon(nS) \rightarrow \bar{d} + X$ decay and their decay widths \cite{CLEO,babar}.
Considering the error bars of the data, with the obtained coupling constant, the variation of $p_0^d$ and $\Lambda^d$ can form a band. The results are plotted in Fig.~\ref{fig:width}. The band within the solid lines in each figure is our calculated results for $\bar{d}$ case. One can see that the experimental distribution can be reasonably described. The momentum of the generated $\bar{d}$ in the $\Upsilon(nS)$ decay is centralized around 1.0 GeV. For $\Upsilon(1S)$, $p_0^d$ and $\Lambda^d$ are  in the ranges $0.3 - 0.4$ GeV and $0.8-1$ GeV. For $\Upsilon(2S)$ and $\Upsilon(3S)$, they are in the ranges $0.2 - 0.3$ GeV and $0.8-0.9$ GeV. With $g^{\text{eff}}=1.0 \times 10^{-3}$, we get the decay width of $\Upsilon(1S)$ $(125 - 165) \times 10^{-5}$ keV, which is close to the experimental data. The effective coupling constant
$g^{\text{eff}}$ is smaller for $\Upsilon(2S)$ and $\Upsilon(3S)$ decay. They are $0.86 \times 10^{-3}$ and $0.5 \times 10^{-3}$ and the corresponding decay widths are in the range $(87 - 105) \times 10^{-5}$ keV and $(39 - 49) \times 10^{-5}$ keV. We should mention that we did not include the momentum dependence of the effective coupling constant, but determine the coupling constant for different $\Upsilon(nS)$ states by fitting their own decay widths.

We have reproduced the experimental data for $\Upsilon(nS)\rightarrow\bar{d}+X$. Now we can make some predictions for the decay $\Upsilon(nS)\rightarrow\bar{d}^\ast(2380)+X$. It is straightforward to use the same parameters and replace the wave function of deuteron with the wave function of $d^\ast$. The obtained momentum distributions of $\bar{d^\ast}$ are also shown in Fig.~\ref{fig:width} with the bands formed by the dotted lines. It can be seen that with the same parameters, the momentum distributions of $\bar{d}$ and $\bar{d^\ast}$ are close to each other. Both of them are centralized around 1 GeV. The widths is about $20\%$ smaller for the $\bar{d^\ast}$ case. For example, the maximal decay widths are 136 $\times 10^{-5}$ keV, 85 $\times 10^{-5}$ keV and 40 $\times 10^{-5}$ keV for $\Upsilon(1S)$, $\Upsilon(2S)$ and $\Upsilon(3S)$, respectively.

In the semi-inclusive decay, theoretically, we do not know exactly the function $F(p_d)$ or $F(p_d^{\ast})$ because of lacking the information of $X$. For this reason, when we make a prediction for the production of $\bar{d^{\ast}}$,
we let the values of $p_0^{d^\ast}$ and $\Lambda^{d^\ast}$ in relatively broad ranges. Since the mass of $\bar{d^\ast}$ is larger than $\bar{d}$, $p_0^{d^{\ast}}$ and $\Lambda^{d^\ast}$ could be smaller than $p_0^d$ and $\Lambda^d$.
For example, the lower limit of $p_0^{d^{\ast}}$ and $\Lambda^{d^\ast}$ are chosen to be 0.1 and 0.3 GeV for $\Upsilon(1S)$ decay. With this choice, the obtained decay width is about one magnitude smaller than that in deuteron case. The momentum distribution of the produced $\bar{d}^\ast$ is shifted to a place peaked around 0.3 GeV. The situation is the same for the $\Upsilon(2S)$ and $\Upsilon(3S)$. The results at lower limit are shown by dashed line on the left of each figure in Fig.~\ref{fig:width}. Therefore, in $d^{\ast}$ case, with the relatively broad ranges of parameters, the obtained widths are $(18-136)\times10^{-5}$ keV, $(7.8-85)\times10^{-5}$ keV and $(5.3-40)\times10^{-5}$ keV for $\Upsilon$ 1S, 2S and 3S states, respectively. Even at the lower limit, $\bar{d^\ast}$ can still be measured by the current experimental facility. This means it is possible to find $\bar{d^{\ast}}$ with momentum $0.3 - 1$ GeV in the semi-inclusive decay of $\Upsilon(nS)$. Finally, we summarize the parameters and the corresponding results in Table \ref{tab1}.

\begin{table}
\begin{center}
\caption{The fitted parameters and the resultant decay widths with the mass parameters: $m_{\bar{b}}=4.96$ GeV, $m_B=5.28$ GeV, $m_q=0.313$ GeV.}
\begin{ruledtabular}
\begin{tabular}{ccccccccccccc}\label{tab1}
State & $g^{\text{eff}} $ & $p_0^{d}$ (GeV) & $\Lambda^{d}$ (GeV) & $\Gamma_{\bar{d}}$ ($10^{-5}$ keV)
& $p_0^{d^*}$ (GeV) & $\Lambda^{d^*}$ (GeV) & $\Gamma_{\bar{d}^\ast}$ ($10^{-5}$ keV) \\
\hline
$1S$ & $1.0\times10^{-3}$ & $0.3-0.4$ & $0.8-1$ & $125-165$ & $0.1-0.4$ & $0.3-1$ & $18-136$   \\

$2S$ & $0.66\times10^{-3}$ & $0.2-0.3$ & $0.8-0.9$ & $87-105$ & $0.1-0.3$ & $0.2-0.9$ & $7.8-85$  \\

$3S$ & $0.5\times10^{-3}$ & $0.2-0.3$ & $0.8-0.9$ & $39-49$ & $0.05-0.3$ & $0.3-0.9$ & $5.3-40$\\
\end{tabular}
\end{ruledtabular}
\end{center}
\end{table}

\section{Summary}\label{section4}

We study the possibility to search $\bar{d}^\ast(2380)$ in the $\Upsilon(nS)$ (n=1, 2, 3) decay in the framework of SU(3) chiral quark model. Utilizing the unitarity of $S$-matrix and crossing symmetry,  the expression for decay widths can be converted to the imaginary part of the forward scattering between $d/d^{\ast}$ and $\Upsilon(nS)$. The processes are governed by the interaction between quarks and $B$ meson. To obtain reasonable result, we firstly study $\Upsilon\rightarrow\bar{d}+X$ channel to fit the unknown parameters. The wave functions of deuteron and $d^\ast$ obtained in our previous work are used. The realistic wave functions of $\Upsilon(nS)$ are obtained by solving the Schr\"{o}dinger equation with a Cornell potential. In the calculation, we examine the effect from both the intermediate pseudoscalar and vector mesons. In fact, due to the momentum $p_{\bar{b}}$ and $p_{q}$ being much smaller than $m_{\bar{b}}$ and $m_{q}$, the contribution of pseudoscalar and vector meson can be summed up resulting in a effective coupling constant $g^{\text{eff}}$. Because of the lack of information on other particles in the final state, we insert a phenomenological Gaussian form factor to describe the momentum distribution of $\bar{d}^\ast$ and $\bar{d}$. We fit the parameters in our model for the deuteron case, and then make some predictions for $\bar{d}^\ast$ case with relatively broad ranges of the parameters accordingly. Our final results show that it is likely to find $\bar{d}^\ast$ in the momentum region $(0.3-1)$ GeV in the $\Upsilon(nS)$ decay. The widths for $\bar{d}^\ast$ in $\Upsilon(nS)$ semi-inclusive decay are about $(18-136)\times10^{-5}$ keV, $(8-85)\times10^{-5}$ keV and $(5-40)\times10^{-5}$ keV for 1S, 2S and 3S states, respectively. We should mention that our results are based on our model calculation. For some other mechanisms, for example, $\bar{d^\ast}$ is produced through gluon processes perturbatively or nonperturbatively, the results might variate in some degree.

\section*{Acknowledgments}
The authors thank F. Huang for providing the wave
functions of the $d^\ast$ and deuteron, and thank Z. X. Zhang,
C. Z. Yuan, H. B. Li and C. P. Shen for helpful discussions. This work is supported in part by the National Natural Science Foundation of China under
Grant Nos. 11475186, 11475192, 11521505, and
11565007, the Sino-German CRC 110 "Symmetries and the Emergence of
Structure in QCD" project by NSFC under the grant No.11621131001,
the Key Research Program of Frontier Sciences, CAS, Grant No.
Y7292610K1, and the IHEP Innovation Fund under the Grant No.
Y4545190Y2.

\end{document}